\documentclass[11pt]{article}
\usepackage{amsfonts,bm,bbm,amssymb,caption}
\usepackage{upgreek} 
\usepackage{mathrsfs}
\usepackage{amsmath}	
\usepackage{cite}
\usepackage{graphicx}
\usepackage{rotating}
\usepackage{hyperref}
\usepackage{verbatim}
\usepackage[all]{xy}

\csname @addtoreset\endcsname{equation}{section}
\newcommand{\beq}{\begin{equation}}
\newcommand{\eeq}{\end{equation}}
\newcommand{\bea}{\begin{eqnarray}}
\newcommand{\eea}{\end{eqnarray}}
\newcommand{\ba}{\begin{array}}
\newcommand{\ea}{\end{array}}
\newcommand{\bit}{\begin{itemize}}
\newcommand{\eit}{\end{itemize}}
\newcommand{\nn}{\nonumber}

\newcommand{\mezzo}{\frac{1}{2}}
\newcommand{\complesso}{{\ \hbox{{\rm I}\kern-.6em\hbox{\bf C}}}}
\newcommand{\reale}{{\hbox{{\rm I}\kern-.2em\hbox{\rm R}}}}

\newcommand{\p}{\partial}

\renewcommand{\a}{\alpha}
\renewcommand{\b}{\beta}

\newcommand{\Er}{{\mathcal{E}}}

\renewcommand{\k}{\kappa}
\renewcommand{\l}{\lambda}
\renewcommand{\L}{\Lambda}

\newcommand{\m}{\mu}

\newcommand{\n}{\nu}
\renewcommand{\r}{\rho}
\newcommand{\s}{\sigma}

\renewcommand{\t}{\theta}

\newcommand{\om}{\omega}

\begin{document}

\begin{titlepage}
\begin{flushright}
CECS-PHY-13/01
\end{flushright}
\vspace{2.5cm}
\begin{center}
\renewcommand{\thefootnote}{\fnsymbol{footnote}}
{\LARGE \bf Embedding hairy black holes in a magnetic universe}
\vskip 4mm
\vskip 30mm
{\large {Marco Astorino\footnote{marco.astorino@gmail.com}}}\\
\renewcommand{\thefootnote}{\arabic{footnote}}
\setcounter{footnote}{0}
\vskip 10mm
{\small \textit{
Centro de Estudios Cient\'{\i}ficos (CECs), Valdivia,\\ 
Chile\\}
}
\end{center}
\vspace{5.2 cm}
\begin{center}
{\bf Abstract}
\end{center}
{Ernst's solution generating technique is adapted to Einstein-Maxwell theory conformally (and minimally) coupled to a scalar field. This integrable system enjoys a SU(2,1) symmetry which enables one to move, by Kinnersley transformations, though the axisymmetric and stationary solution space, building an infinite tower of physically inequivalent solutions. \\
As a specific application, metrics associated to scalar hairy black holes, such as the ones discovered by Bocharova, Bronnikov, Melnikov and Bekenstein, are embedded in the external magnetic field of the Melvin universe. 
}
\end{titlepage}

\section{Introduction}

Exact solutions in the realm of general relativity are of immense interest and utility but, because of the non-linear behaviour of the theory, they are not easy to discover. For this scope some very powerful solution generating techniques were built in the last decades, relying basically on the integrability properties of the system. The most famous branches in this field are the Ernst's \cite{ernst1}-\cite{ernst-complex} and the Belinsky-Zakharov approaches \cite{belinski} for stationary axisymmetrical space-times in the Einstein-Maxwell theory of gravity. These techniques not only are a useful tool to construct non-trivial solutions such as black holes in magnetic universes \cite{ernst-magnetic},\cite{kerr-magnetic} or rotating multi-black hole solutions, but also were fundamental in the proof, by Helers and Ernst, of the Gerosh conjecture  which states that, in principle, $all$ electro-vacuum stationary, axisymmetric, spinning mass solutions could be generated by one particular solution (e.g. Minkowski space-time) by means of an infinite sequence of transformations of a certain group\cite{hauser-ernst},\cite{ernst3}. All these approaches are strongly theory dependent and it is difficult to apply them even for small modifications of the theory's action. For instance just the addition of the cosmological constant makes this method hard to generalise, as can be seen in \cite{asto-ernst-lambda} (in this case the problem is related to a reduction of symmetry of the moduli space which makes the system not explicitly integrable any more).\\
Here we are interested in extending the Ernst's technique for Einstein-Maxwell theory to the presence of a minimally and conformally coupled scalar field. In this case the integrability property is preserved so the Ernst approach can be directly extended, as can be seen in section \ref{ernst-minimally} and \ref{ernst-conformally}.\\
Besides the fact that the literature for gravitational systems coupled with scalar field is wide both from a theoretical and phenomenological point of view, actual astrophysical support for this kind of matter is not proven. Cosmologist use scalar fields in some models of inflation or employ them to describe dynamical models for dark matter or dark energy. Neither are fundamental scalar fields known in nature, apart some recent footprint of the Higgs field found at CERN, which is anyway of a different kind than the ones considered here. Nevertheless the theoretical interest for those conformally coupled scalar field has arisen, at least since the seventies, when Bekenstein made use of it to find the first counter-example to the famous Wheeler's \emph{``Black holes have not hairs''} conjecture. For an historical perspective see \cite{bekenstein20}. In fact this matter is at least viable from a theoretical point of view in the sense that it does not violate most of the energy conditions, so if it is not endorsed at the moment by observation, it is at least plausible, possibly just at an effective level. \\
The black hole solution for general relativity conformally coupled with a scalar filed was first found by Bocharova, Bronnikov and Melnikov in \cite{BBM} and then independently studied by Bekenstein in \cite{bekenstein1} and \cite{bekenstein2} (henceforth we will call this metric BBMB). It is a static solution of Einstein-Maxwell theory, whose stationary rotating generalisation is not known. The formalism developed in this paper could be of some utility in this direction or in other generalizations of the BBMB black hole as well, for instance embedding it in a external magnetic field as was done by Ernst  in \cite{ernst-magnetic} for the Schwarzschild and Reissner-Nordstrom black holes by means of a Harrison transformation. This point is addressed in section \ref{ernst-conformally}. Black holes embedded in a external magnetic source, such as the one of the Melvin universe, are of some astrophysical interest because, especially at the center of galaxies, currents in the accretion disk around a black hole can likely generate such kind of magnetic fields.

\section{Ernst's solution generating technique for Einstein-Maxwell theory with a minimally coupled scalar field}
\label{ernst-minimally}

\subsection{Equations of motion}
\label{ersnt-min-eq}

Consider the action for general relativity coupled to the Maxwell electromagnetic field and to a minimally coupled scalar field $\Psi$:
\beq  \label{minimal-action}
                       I[g_{\m\n}, A_\m, \Psi] =  \frac{1}{16 \pi G}  \int d^4x  \sqrt{-g} \left[ \textrm{R} - \frac{G}{\m_0}F_{\m\n} F^{\m\n} - \k \ \nabla_\m \Psi \nabla^\m \Psi \right]  \ \ \ .
\eeq

The gravitational, electromagnetic and scalar field equations are obtained by extremising with respect to metric $g_{\m\n}$, the electromagnetic potential $A_\m$ and the scalar field $\Psi$ respectively: 
\bea  \label{min-field-eq}
                        &&   \textrm{R}_{\m\n} -   \frac{\textrm{R}}{2}  g_{\m\n} = \frac{2G}{\m_0} \left( F_{\m\r}F_\n^{\ \r} - \frac{1}{4} g_{\m\n} F_{\r\s} F^{\r\s} \right) + \k \left( \p_\m \Psi \p_\n \Psi - \mezzo g_{\m\n} \p_\s \Psi \p^\s \Psi \right)  \quad ,       \\
                        &&   \partial_\m ( \sqrt{-g} F^{\m\n}) = 0  \ \quad , \\
                        &&   \Box \Psi=0 \quad .
\eea
We are interested in stationary and axisymmetric space-times characterized by two commuting killing vectors $\p_t,\p_\varphi$ that, for this minimal coupling, are given, in the most general way\footnote{As explained in \cite{heusler-book}, section 3.4.}, by the Lewis-Weyl-Papapetrou metric: 
\beq \label{min-axis-metric}
                         ds^2 =  f \left( dt - \om d\varphi \right)^2 - f^{-1} \left[ r^2 d\varphi^2 + e^{2\gamma}  \left( d r^2 + d z^2 \right) \right] \ ,
\eeq
where all the functions $f,\omega,\gamma$ depend just on the coordinates $(r, z)$ and $\k=8\pi G$. The most generic electromagnetic potential and scalar field compatible with this symmetry can be written as $A=A_t(r, z) dt + A_\varphi(r,z) d\varphi$ and $\Psi(r,z)$ respectively.\\
In terms of the form of the metric (\ref{min-axis-metric}) the principal gravitational field equations ($GE^\m_{\ \n}$) are\footnote{To match the standard Ernst notation $G/\mu_0$ can be normalized to 1 without loss of generality. Any signs discrepancy respect reference \cite{ernst2} are due to several renowned typos of the latter, as admitted in \cite{ernst-complex}.}
\bea \label{min-grav-field-eq}
          GE^\varphi_{\ t}:        &  \overrightarrow{\nabla} \cdot \left[ r^{-2} f^2 \overrightarrow{\nabla} \omega - 4 \frac{G}{\m_0} r^{-2} f A_t (\overrightarrow{\nabla} A_\varphi + \omega \overrightarrow{\nabla} A_t) \right] = 0 \qquad ,&\\
\hspace{-0.2cm}         GE^t_{\ \varphi} - GE^\varphi_{\ \varphi}:      &  f \nabla^2 f = (\overrightarrow{\nabla} f)^2 - r^{-2} f^4 (\overrightarrow{\nabla} \omega)^2 + 2\frac{G}{\m_0}f \left[(\overrightarrow{\nabla} A_t)^2 + r^{-2} f^2 (\overrightarrow{\nabla} A_\varphi + \omega \overrightarrow{\nabla} A_t)^2 \right] ; \ \  \  &
\eea
while the Maxwell ($ME^\m$) and scalar ($SE$) field  equations become:
\bea  \label{min-max-field-eq}
      ME^t: \qquad \ \  &   \overrightarrow{\nabla}  \cdot \left[ f^{-1} \overrightarrow{\nabla} A_t - r^{-2} f \omega (\overrightarrow{\nabla} A_\varphi + \omega \overrightarrow{\nabla} A_t) \right]=0 \qquad , & \\
      ME^\varphi:   \qquad \ \             &      \overrightarrow{\nabla} \cdot \left[  r^{-2} f (\overrightarrow{\nabla} A_\varphi + \omega \overrightarrow{\nabla} A_t)  \right] = 0 \qquad , &  \\
    \label{se}  SE:       \qquad   \ \      &       \nabla ^2 \Psi = 0  \qquad . &
\eea
The differential vectorial operators appearing here are the standard flat ones in polar cylindrical coordinates. As can be seen the scalar field remains decoupled from the gravitational ($GE$) and electromagnetic ($ME$) equations and the $\gamma$ does not appear, so it can be obtained by quadrature after having detected the other functions. This set of equations (\ref{min-grav-field-eq}) - (\ref{se}) can be reduced to two complex and one real equations as follows:
\bea 
     \label{ee-ernst}  \left( \text{Re} \ \Er + | \mathbf{\Phi} |^2 \right) \nabla^2 \Er   &=&   \left( \overrightarrow{\nabla} \Er + 2 \ \mathbf{\Phi^*} \overrightarrow{\nabla} \mathbf{\Phi} \right) \cdot \overrightarrow{\nabla} \Er   \quad ,       \\
     \label{em-ernst}   \left( \text{Re} \ \Er + | \mathbf{\Phi} |^2 \right) \nabla^2 \mathbf{\Phi}  &=& \left( \overrightarrow{\nabla} \Er + 2 \ \mathbf{\Phi^*} \overrightarrow{\nabla} \mathbf{\Phi} \right) \cdot \overrightarrow{\nabla} \mathbf{\Phi} \quad , \\
     \label{sclr-ernst}     \nabla ^2 \Psi &=& 0  \qquad . 
\eea
taking advance of the system's integrability and introducing two complex ($\Er,\Psi$) fields such that:
\beq \label{def-psi-Er} 
       \mathbf{\Phi} := A_t + i \tilde{A}_\varphi  \qquad , \qquad \qquad     \Er := f - |\mathbf{\Phi} \mathbf{\Phi}^*| + i h  \qquad ,
\eeq
where $\tilde{A}_\varphi$ and $h$ are defined as:
\bea
    \label{A-tilde} \overrightarrow{\nabla} \tilde{A}_\varphi &:=& - f r^{-1} \overrightarrow{e}_\varphi \times (\overrightarrow{\nabla} A_\varphi + \omega  \overrightarrow{\nabla} A_t ) \\
    \label{h}    \overrightarrow{\nabla} h &:=& - f^2 r^{-1} \overrightarrow{e}_\varphi \times \overrightarrow{\nabla} \omega - 2 \ \textrm{Im} (\mathbf{\Phi}^*\overrightarrow{\nabla} \mathbf{\Phi} )  \qquad .
\eea

Remarkably enough these equations of motion (\ref{ee-ernst})-(\ref{sclr-ernst}) can be derived by an effective action principle: 
\beq \label{maxw-GR+MinScl-action}
     S [\Er,\mathbf{\Phi},\Psi] = \int r d r d z d \varphi \left[ \frac{(\overrightarrow{\nabla}\Er + 2 \mathbf{\Phi^*} \overrightarrow{\nabla} \mathbf{\Phi} ) \cdot (\overrightarrow{\nabla}\Er^* + 2 \mathbf{\Phi} \overrightarrow{\nabla} \mathbf{\Phi}^* )}{(\Er + \Er^* + 2 \mathbf{\Phi} \mathbf{\Phi}^*)^2} - \frac{2 \overrightarrow{\nabla} \mathbf{\Phi} \cdot \overrightarrow{\nabla} \mathbf{\Phi}^*}{\Er + \Er^* + 2 \mathbf{\Phi} \mathbf{\Phi}^*} -\frac{\k}{2} \overrightarrow{\nabla} \Psi \cdot \overrightarrow{\nabla} \Psi \right] 
\eeq
The homothetic symmetries of the action (\ref{maxw-GR+MinScl-action}) are those that leave the equations of motion (\ref{ee-ernst}- \ref{sclr-ernst}) invariant. They form a nine real parameters group $SU(2,1) \times U(1)$ represented by these finite transformations:
\bea 
      I)    && \Er \longrightarrow \Er' = \l \l^* \Er  \qquad \ \quad \qquad \ ,  \quad \mathbf{\Phi} \longrightarrow  \mathbf{\Phi}' = \l \mathbf{\Phi} \qquad \qquad \qquad \quad , \ \ \ \Psi \longrightarrow \Psi' = \Psi \nn \\
      II)   && \Er \longrightarrow \Er' = \Er + i \ b \qquad  \ \ \ \qquad, \quad \mathbf{\Phi} \longrightarrow  \mathbf{\Phi}' = \mathbf{\Phi} \qquad \qquad \qquad \quad \ , \ \ \ \Psi \longrightarrow \Psi' = \Psi  \nn \\
      III)  && \Er \longrightarrow \Er' = \Er/(1+ic\Er) \qquad \quad , \quad  \mathbf{\Phi} \longrightarrow  \mathbf{\Phi}' = \mathbf{\Phi}/(1+ic\Er) \qquad \ \ \ , \ \ \ \Psi \longrightarrow \Psi' = \Psi \nn \\
      IV)   && \Er \longrightarrow \Er' = \Er - 2\b^*\mathbf{\Phi} - \b\b^* \ \ \ , \quad \mathbf{\Phi} \longrightarrow  \mathbf{\Phi}' = \mathbf{\Phi} + \b \qquad \qquad \ \quad , \ \ \ \Psi \longrightarrow \Psi' = \Psi \nn \\
      V)    && \Er \longrightarrow \Er' = \frac{\Er}{1-2\a^*\mathbf{\Phi}-\a\a^*\Er} \  , \ \ \ \mathbf{\Phi} \longrightarrow  \mathbf{\Phi}' = \frac{\mathbf{\Phi}+\a\Er}{1-2\a^*\mathbf{\Phi}-\a\a^*\Er}\ , \ \ \Psi \longrightarrow \Psi' = \Psi \nn \\ 
      VI)   &&  \Er \longrightarrow \Er' = \Er  \qquad \qquad \qquad \quad \ \ , \quad \mathbf{\Phi} \longrightarrow  \mathbf{\Phi}' = \mathbf{\Phi}  \qquad \qquad \qquad \quad \ , \ \ \Psi \longrightarrow \Psi' = \Psi + d \nn  
\eea
where $b, c, d \in \mathbb{R}$ and $\a, \l,\b \in \mathbb{C}$. More generally, instead of the last term in the action (\ref{maxw-GR+MinScl-action}), it is possible to have a sigma model for a collection of scalar fields $\Psi_A$:
$\frac{\kappa}{2}G_{AB} \overrightarrow{\nabla} \Psi^A  \cdot \overrightarrow{\nabla} \Psi^B $, as done without the electromagnetic field in \cite{heusler-book} or \cite{heusler-art}. In this case the group of symmetry is, at least, $SU(2,1) \times \mathcal{G}$, where $\mathcal{G}$ is the group of homothetic symmetry of the scalar matter. The case we will consider here is just the simplest: $G_{AB}=1$. \\
$(I-V)$ are the standard $SU(2,1)$ Kinnersley symmetries, while $(VI)$ is just a trivial shift $U(1)$. Some of these transformations physically represent gauge transformations, that is they can be reabsorbed by some diffeomorphism of the resulting metric, while some of them give inequivalent space-times being in fact able to change the charges, the asymptotic behaviour, the electromagnetic field content, etc. So the effective group of transformation is actually smaller than $SU(2,1)$. \\
In principle we suspect that any axisymmetric metric of the Einstein-Maxwell theory minimally coupled with a scalar field could be obtained, from a fixed seed, by means of subsequent transformations $(I-VI)$. The case with a vanishing scalar field was proven by Hauser and Ernst in \cite{hauser-ernst}. In practice it is not easy to find this sequence and moreover not all transformations preserve the asymptotic behaviour of the previous solution. In particular, in the next section, we will mostly be interested in the Harrison transformation $(V)$, which is well known to enable one to embed one's favourite asymptotically flat space-times in a magnetic universe. Ernst was able to embed Schwarzschild and the whole Kerr-Newman family of black holes into the Melvin magnetic universe \cite{ernst-magnetic}, \cite{kerr-magnetic}. Note that, after being immersed in the external magnetic field, these black hole solution are not of type D in the Petrov classification any more.
\\

\subsection{Magnetising the Fisher, Janis, Robinson and Winicour solution}
\label{magne-jrw}

Here we take advantage of the formalism of section \ref{ersnt-min-eq} to embed the solution of Fisher and Janis, Robinson, Winicour (henceforth FJRW) \cite{Fisher},\cite{janis} in a external magnetic field. That metric describes a static, asymptotically flat solution for Einstein gravity minimally coupled with a scalar field. Since it is plagued by some non-physical features, it is not considered of physical interest, but our strategy is to use it as an intermediate step towards the more physical BBMB black hole family. The metric and the associated scalar field read:
\bea \label{janis}
     ds^2 &=&  -\left(1-\frac{2m}{R}\right)^A d\tau^2 + \frac{d R^2}{\left(1-\frac{2m}{R}\right)^A} + \left(1-\frac{2m}{R}\right)^{1-A} R^2 \big(d\theta^2 + \sin^2\theta \ d\phi^2 \big) \\
    \label{psi-min} \Psi &=& \sqrt{\frac{1-A^2}{2 \kappa}} \log \left(1-\frac{2m}{R}\right)     
\eea
where the real parameter $A\in[0,1]$. For $A\in[0,1)$ the surface $R=2m$ has the Ricci squared curvature invariant $\textrm{R}_{\m\n} \textrm{R}^{\m\n}$ unbounded so it is a naked singularity, while for $A=1$ it is evident that we have the Schwarzschild Black hole. Another interesting value is $A=1/2$ because, though it is an non-physical solution in this minimal frame, it can be used, via the Bekenstein technique (which basically consists in a conformal rescaling), to obtain the BBMB black hole in the conformal frame. For this reason, henceforward in this section, the parameter $A$ will be fixed to $1/2$. The case with generic $A$ for the magnetized FJRW can be extracted in the next section \ref{magnetic-penney}, just by setting $e_0=0$ (or equivalently $b=0$) in the metric (\ref{magn-penney}). \\
Now we want to embed this solution, which will be considered as our seed metric, in the Melvin magnetic universe. In absence of the scalar field the standard procedure consists in using the Harrison transformation $(V)$, so we will do the same. For this purpose is more conventional (with respect to the standard literature \cite{ernst-magnetic}-\cite{kerr-magnetic}) to use another form of the Weyl-Lewis-Papapetrou metric (\ref{min-axis-metric}) obtained from this latter by double Wick rotation $(t,\varphi) \rightarrow (i\phi,i\tau)$:
\beq \label{min-axis-metric-wick}
                         ds^2 =  - f \left( d\phi - \om d \tau \right)^2 + f^{-1} \left[ r^2 d \tau^2 - e^{2\gamma}  \left( d r^2 + dz^2 \right) \right] \ ,
\eeq
Note that after the Wick rotation the electromagnetic complex potential (\ref{def-psi-Er}) become $\mathbf{\Phi}=A_\phi + i \tilde{A}_\tau$.
Comparing (\ref{janis}) and (\ref{min-axis-metric-wick}) we get the complex seed potential associated with the killing vector $\p_\phi$:
\beq
        \mathbf{\Phi}_0 = 0           \qquad  ,  \quad  \qquad             \Er_0 = f_0 = - \sqrt{R^4-2mR^3} \sin^2 \theta \quad .
\eeq
Then we apply the Harrison transformation to get:
\beq
        \mathbf{\Phi} = \frac{B}{2} \frac{\Er_0}{\Lambda}      \qquad  ,   \quad     \qquad       \Er = \frac{\Er_0}{\Lambda} 
\eeq
where we call\footnote{In Ernst's notation $\a=-\frac{B_0}{2}$ \cite{ernst-magnetic}, which imply also a switch of the first minus sign in (\ref{D_omega}).}:
$$ \a=\frac{B}{2}  \qquad \qquad  \Lambda = 1 - \a\a^* \Er_0 = 1 + \frac{B^2}{4} \sqrt{R^4-2mR^3} \sin^2\theta \quad .$$
So the magnetised Janis-Robinson-Winicour space-times becomes:
\beq \hspace{-0.2cm} \label{magn-jrw}
      ds^2 =  \Lambda^2 \left(  -\sqrt{1-\frac{2m}{R}} \ d\tau^2 + \frac{d R^2}{\sqrt{1-\frac{2m}{R}}} +  \sqrt{R^4 - 2mR^3} \ d \theta^2 \right) + \frac{\sqrt{R^4-2mR^3} \sin^2 \theta}{\Lambda^2} \ d\phi^2 \  \ ;
\eeq 
the scalar field remains unchanged as in (\ref{psi-min}) while the magnetic field is given by:
\beq  \label{A_jrw}
          A_\phi=\mathbf{\Phi}= -\frac{B}{2} \frac{\sqrt{R^4-2mR^3} \sin^2 \theta}{1 + \frac{B^2}{4} \sqrt{R^4-2mR^3} \sin^2\theta}
\eeq
This solution still contains non-physical features, such as naked singularities, as the non-magnetic one, so it will be considered only a mathematical step towards a less pathological space-time that will be analysed in section \ref{magnetic-BBMB}. 

\subsection{Magnetising the Penney solution}
\label{magnetic-penney}

Penney in \cite{penney} has found, for the Einstein-Maxwell theory minimally coupled to a scalar field, a generalisation of the FJRW metric in presence of a non-null electric field. For our purposes it is best expressed as follows: 
\bea \label{penney-psi} 
      \Psi&=&\sqrt{\frac{1-A^2}{2\kappa}} \log \left( \frac{R-a}{R-b} \right)   \\
     \label{penney-Atau} A_\tau &=& \frac{(b-a)(R-a)^A}{ b (R-a)^A -a (R-b)^A } \sqrt{\frac{b}{a} \frac{\m_0}{G}}  \\
  \label{penney-metric}    ds^2 &=& -\textrm{e}^{-\a} d\tau^2 + \textrm{e}^\a dR^2 + \textrm{e}^\b \big(d\t^2 + \sin^2 \t \ d\phi^2 \big) \qquad ,
\eea
where
\bea 
     \textrm{e}^\a &=& \frac{\left[ b(R-a)^A-a(R-b)^A\right]^2}{(b-a)^2\left[(R-a)(R-b)  \right]^A}   \\
   \label{beta}  \textrm{e}^\b &=& e^\a (R-a)(R-b) \qquad . 
\eea
The $a$ and $b$ real parameters are related to the standard electric charge parameter $e_0$ and mass parameter $m$ in this way: $2m=a+b$ and $ a b = e_0^2 G / A^2 \m_0 $ (again the ratio $G/\m_0$ may be thought to be normalized to 1, without loss of generality, to match the standard Ernst notation). $A$ is a constant parameter belonging to the real interval [0,1], as in  section \ref{magne-jrw}. When $b=0$ the FJRW solution (\ref{janis} - \ref{psi-min}) is retrieved. When $A \in [0,1)$ the Penney solution displays naked singularities but for A=1 it is physically meaningful, in fact it collapses into the Reissner-Nordstrom solution. When $A=1/2$ it can be shifted into the conformal frame, by a conformal transformation, giving the charged BBMB metric. For this reason (\ref{penney-metric}) represent a good seed to obtain a magnetised charged black hole in the conformal frame as will be done in section \ref{Q-BBMB-magn}.\\
Comparing the Penney metric (\ref{penney-metric}) with to the Weyl-Lewis-Papapetrou (\ref{min-axis-metric-wick}), according to definitions (\ref{def-psi-Er}), we can extract:
\bea 
         r &=& \sqrt{(R-a)(R-b)} \sin \t \\
       f_0 &=& - \textrm{e}^\b \sin^2 \t  \\
         \mathbf{\Phi}_0&=&\tilde{A}_{\tau 0} = - i A \sqrt{ab} \cos \t   \\
         \Er_0 &=& - \textrm{e}^\b \sin^2 \t - e_0^2 \cos^2 \t 
\eea
Now it is possible to apply the Harrison transformation $(V)$ (with $\a=B/2$) to magnetise the solution (\ref{penney-psi}-\ref{penney-metric}):
\bea
     \Er &=& \frac{\Er_0}{1-B\mathbf{\Phi}_0 - \frac{B^2}{4}\Er_0} = \frac{-\textrm{e}^\b \sin^2 \t -e_0^2 \cos^2\t}{1+iBe_0\cos\t+\frac{B^2}{4}(\textrm{e}^\b \sin^2 \t +e_0^2 \cos^2\t)} \\
     \mathbf{\Phi} &=& \frac{\mathbf{\Phi}_0 + \frac{B}{2} \Er_0}{1-B\mathbf{\Phi}_0 - \frac{B^2}{4}\Er_0} = \frac{-ie_0\cos\t -\frac{B}{2} (\textrm{e}^\b \sin^2 \t + e_0^2 \cos^2\t)}{1+iBe_0\cos\t+\frac{B^2}{4}(\textrm{e}^\b \sin^2 \t +e_0^2 \cos^2\t)}
\eea
This represents the Penney solution embedded in an external magnetic field, written in terms of the Ernst complex potentials. In case one wants to express it in terms of the more familiar metric, electromagnetic and scalar field, it is sufficient to apply definitions   (\ref{def-psi-Er}) - (\ref{h}):
\bea \label{A_phi}  A_\phi &=& \textrm{Re}(\mathbf{\Phi}) =\frac{- \frac{B}{2} (\textrm{e}^\b \sin^2 \t + e_0^2 \cos \t) - \frac{B^2}{8} (\textrm{e}^\b \sin^2 \t + e_0^2 \cos \t)^2 - B e_0^2 \cos^2 \t}{\left[ 1 + \frac{B^2}{4} (\textrm{e}^\b \sin^2 \t +e_0^2 \cos^2\t) \right]^2 + B^2 e_0^2 \cos^2 \t}  \\
      \tilde{A}_\tau &=& \textrm{Im}(\mathbf{\Phi}) = -e_0 \cos \t \frac{ 1 - \frac{B^2}{4} (\textrm{e}^\b \sin^2 \t + e_0^2 \cos \t)}{\left[ 1 + \frac{B^2}{4} (\textrm{e}^\b \sin^2 \t +e_0^2 \cos^2\t) \right]^2 + B^2 e_0^2 \cos^2 \t} \qquad ,\\
      f &=& \textrm{Re}(\Er) + \mathbf{\Phi} \mathbf{\Phi}^* = \frac{-\textrm{e}^\b \sin^2 \t}{\left[ 1 + \frac{B^2}{4} (\textrm{e}^\b \sin^2 \t +e_0^2 \cos^2\t) \right]^2 + B^2 e_0^2 \cos^2 \t}   \qquad .
\eea
The last unknown metric function $\omega$ can be found, for this particular Harrison transformation (V), thanks to the relation:
\beq \label{D_omega}
  \overrightarrow{\nabla} \omega = \Lambda \Lambda^* \overrightarrow{\nabla} \om_0 - \frac{r}{f_0} (\Lambda^*\overrightarrow{\nabla}\Lambda-\Lambda\overrightarrow{\nabla}\Lambda^*)  \quad ,   
\eeq
where in this case $\omega_0=0$, because the seed (\ref{penney-Atau}-\ref{beta}) we have begun with is static and where $ \Lambda(R,\theta)=1+iBe_0\cos\t+\frac{B^2}{4}(\textrm{e}^\b \sin^2 \t +e_0^2 \cos^2\t) $. In these coordinates the differential operator $\overrightarrow{\nabla}$ can be taken as follows:
$$ \overrightarrow{\nabla} f(R,\t) = \overrightarrow{e}_R \ \sqrt{(R-a)(R-b)} \ \frac{\p f(R,\t)}{\p R} + \overrightarrow{e}_\t \ \frac{\p f(R,\t)}{\p \t} $$
 Thus (\ref{D_omega}) gives:
 \begin{equation*}
\left\{
\begin{array}{rl}
 &     \p_R \om = - e_0 \frac{B^3}{2} (1+\cos^2\t) - e_0 \frac{B}{2} (4-B^2 e_0^2\cos^2\t) \textrm{e}^{-\b} \\
 &   \ \p_\t \omega = e_0  \frac{B^3}{2}  (R-a)(R-b) \ \frac{d\b(R)}{dR} \ \sin\t \cos\t  \qquad \quad . \\
\end{array} \right.
\end{equation*}
The latter equation can be integrated up to a arbitrary function $F(R)$, which can be found from the first.
\bea
      \omega(R,\theta) &=&  \frac{e_0B^3}{4} \sin^2\t \left[ 2A \frac{b(R-a)^A(R-b)-a(R-b)^A(R-a)}{b(R-a)^A-a(R-b)^A} + (1-A) (2R-a-b) \right] +F(R) \nn \\
       F(R) &=& -B^3 e_0 R + (4Be_0-B^3e_0^3) \frac{1}{2Aa} \frac{(a-b)(R-a)^A}{b(R-a)^A-a(R-b)^A} +F_0     \qquad \quad .  \nn
\eea
The magnetised Penney metric, thus, takes the final form:
\beq \label{magn-penney}
                             ds^2=|\Lambda(R,\t) |^2 \left[ -e^{-\a(R)} d\tau^2 + e^{\a(R)} dR^2 + e^{\b(R)} d\t^2  \right] +  \frac{e^{\b(R)} \sin^2\t}{|\Lambda(R,\t) |^2} \left[d\phi-\omega(R,\t) dt\right]^2   
\eeq
The electric potential component  $A_\tau$ follows from the double wick rotated (\ref{A-tilde}):

\beq
\label{A-tilde-dwr} \overrightarrow{\nabla} \tilde{A}_\tau := - f r^{-1} \overrightarrow{e}_\phi \times (\overrightarrow{\nabla} A_\tau + \omega \overrightarrow{\nabla} A_\phi ) 
\eeq
which can be reduced to:
\begin{equation*}
\left\{
\begin{array}{rl}
 &     \p_R (A_\tau+\omega A_\phi)  = \frac{|\Lambda|^2}{e^\beta \sin \theta} \p_\theta \tilde{A}_\tau + A_\phi \p_R \omega  \\
 &    \p_\theta (A_\tau+\omega A_\phi) = -\frac{|\Lambda|^2}{e^\beta \sin \theta} (R-a)(R-b) \p_R\tilde{A}_\tau + A_\phi \p_\theta \omega \quad \qquad .\\
\end{array} \right.
\end{equation*}
Finally the electric potential becomes:
\bea \label{A_tau}
        A_\tau(R,\t) &=& -\frac{3}{8} e_0 B^2 \sin^2 \t \left[ 2 A \frac{b(R-a)^A(R-b)-a(R-b)^A(R-a)}{b(R-a)^A-a(R-b)^A} + (1-A) (2R-a-b)  \right]    \nn \\ 
               &+& \frac{3}{2} B^2 e_0 R + \left( \frac{3}{4} e_0 B^2 A b - \frac{e_0}{aA} \right)  \frac{(a-b)(R-a)^A}{b(R-a)^A-a(R-b)^A}
- \omega A_\phi + \textrm{cost.}
 \eea
In the next section we will combine these outcomes with the Bekenstein technique, in presence of a conformal scalar field, to embed scalar hairy black holes of the BBMB type in an external magnetic field background.  \\

\section{Einstein-Maxwell with a conformally coupled scalar field}
\label{ernst-conformally}

\subsection{BBMB black hole in Melvin magnetic universe}
\label{magnetic-BBMB}

When the scalar filed is conformally coupled to the Einstein-Maxwell theory the action becomes\footnote{An extra conformally invariant potential term, such as $\alpha \hat{\Psi}^4$, might be included to the action (\ref{conformal-action}), but we prefer not consider it here, because it would imply a potential term, in the minimally coupled system (\ref{minimal-action}), which spoils the integrability, and because is not necessary in the BBMB solutions that we will treat.}
\beq  \label{conformal-action}
                       \hat{I}[\hat{g}_{\m\n}, \hat{A}_\m, \hat{\Psi}] =  \frac{1}{16 \pi G}  \int d^4x  \sqrt{-\hat{g}} \left[ \hat{\textrm{R}} - \hat{F}_{\m\n} \hat{F}^{\m\n} - \k \ \left( \nabla_\m \hat{\Psi} \nabla^\m \hat{\Psi}  + \frac{\hat{\textrm{R}}}{6} \hat{\Psi}^2   \right) \right]  \ \ \ .
\eeq
We will denote all the quantities in this conformal frame with a hat: $\hat{g}_{\m\n}, \hat{A}_\m , \hat{\Psi}, \dots$

It was discovered by Bekenstein in \cite{bekenstein1} that a solution (${g_{\m\n},A_\m,\Psi}$) of Einstein-Maxwell gravity minimally coupled to a scalar field can be mapped to a solution ($\hat{g}_{\m\n}, \hat{A}_\m, \hat{\Psi}$) of the Einstein-Maxwell theory with a conformally coupled scalar field (\ref{conformal-action}) by this set of transformations:
\bea 
    \label{conf-tras-psi} \Psi &\longrightarrow & \hat{\Psi}=\sqrt{\frac{6}{\kappa}} \tanh \left(\sqrt{\frac{\kappa}{6}}\Psi \right) \\
    \label{conf-tras-A} A_\m &\longrightarrow & \hat{A}_\m  = A_\m  \\
    \label{conf-tras-g} g_{\m\n} &\longrightarrow & \hat{g}_{\m \n} = \left(1-\frac{\kappa}{6} \hat{\Psi}^2\right)^{-1} g_{\m\n} 
\eea
Actually the original BBMB solution can be obtained by this technique from the $A=1/2$ Fisher, Janis, Robinson and Winicour one (\ref{janis}). So we play the same game starting with the magnetised FJRW solution (\ref{magn-jrw}),(\ref{A_jrw}),(\ref{psi-min}) and applying the transformations (\ref{conf-tras-psi})-(\ref{conf-tras-g}) to pass to the conformal frame.
After a coordinate transformation in the radial coordinate $R \rightarrow \rho/(1-\frac{m}{2\rho})$, we obtain:

\bea
 &&  \label{magn-scalar} \hat{\Psi}=\sqrt{\frac{6}{\kappa}} \left( 1 - \frac{2\rho}{m} \right)^{-1}  \\
  &&   A_\phi = -\frac{2 B}{\Lambda} \rho^3 \frac{\rho -m}{(2\rho-m)^2}\sin^2 \theta   \\
   && \label{magn-BBMB} \hat{ds}^2 = \Lambda^2 \left[ - \left(  1 - \frac{m}{2\rho}  \right)^2 d\tau^2 + \frac{d\rho^2}{\left(  1 - \frac{m}{2\rho}  \right)^2} + \rho^2 d\theta^2 \right] + \frac{\rho^2 \sin^2 \theta}{\Lambda^2} d\phi^2 \qquad ,
\eea
where
\beq
      \Lambda(\rho,\theta)=1+B^2\rho^3\frac{\rho -m}{(2\rho-m)^2}\sin^2 \theta  \qquad \ \ .
\eeq
\\
This solution represents a BBMB black hole embedded in the Melvin magnetic universe. In fact, as can be easily seen from the limit of the mass parameter $m \rightarrow 0$, the Melvin universe is exactly recovered:
\bea
      A_\phi &=& -\frac{B^2}{2} \frac{\rho^2 \sin^2 \t}{1+\frac{B^2}{4} \rho^2 \sin^2 \t}  \quad , \qquad \Psi=0 \\
      ds^2&=&\left( 1+\frac{B^2}{4} \rho^2 \sin^2 \t \right)^2 \left[ -d\tau^2 + d\rho^2 +\rho^2 d\t^2 \right] + \frac{\rho^2 \sin^2 \t}{\left(1+\frac{B^2}{4} \rho^2 \sin^2 \t \right)^2} d\phi^2
\eea
The magnetic universe found by Melvin is a static, non-singular, cylindrical symmetric space-time in which there exists an axial magnetic field aligned  with the z-axis. It describes a universe containing a parallel bundle of electromagnetic flux held  together by its own gravitational field. 
Actually this magnetic universe mimics also the asymptotic behaviour (for large $\rho$) of the metric (\ref{magn-BBMB}). \\
While the limit of vanishing external magnetic field ($B\rightarrow0$) of the solution (\ref{magn-scalar})-(\ref{magn-BBMB}) gives, as expected, the BBMB black hole:

\bea
 &&  \hat{\Psi}=\sqrt{\frac{6}{\kappa}} \left( 1 - \frac{2\rho}{m} \right)^{-1}  \\
   && \label{BBMB} \hat{ds}^2 =   - \left(  1 - \frac{m}{2\rho}  \right)^2 d\tau^2 + \frac{d\rho^2}{\left(  1 - \frac{m}{2\rho}  \right)^2} + \rho^2 \big( d\theta^2 + \sin^2 \theta d\phi^2 \big) 
\eea
The global causal structure of of the magnetised black holes is generally very close to their non-magnetised relatives, since any slice of constant $\phi$ gives a three-dimensional space-time whose metric coincides with the non-magnetised metric multiplied by a conformal factor that does not deform the casual structure. So the magnetised solutions share with standard black holes (i.e. B=0) the same radial null geodesics, event horizons and trapped surfaces. In this particular case of BBMB black holes the analysis might be more subtle because the conformal factor $\L(\rho,\t)$ appears to be divergent on the surface $\rho=m/2$, in this set of coordinates. Anyway the electric potential $A_\phi$ remains finite everywhere.  The curvature invariants such as $ \textrm{R}^{\m\n} \textrm{R}_{\m\n}$ or $\textrm{R}^{\m\n\s\l}\textrm{R}_{\m\n\s\l}$ show divergences for $\rho=0$, as usual for BBMB black holes, but also on the poles ($\theta=0, \pi$) of the surface $\rho=m/2$\footnote{At least reaching the poles along some particular directions.}. This surface constitutes the event horizon of the BBMB black hole, where it is well known that also the scalar field of that solution is divergent\footnote{As widely discussed by de Witt and Bekenstein in \cite{bekenstein2}: "the infinity in the scalar field is not physically pathological because it is not associated to a infinite potential barrier for test scalar charges, it does not cause the termination of any trajectory of these test particles at finite proper time and it is not connected with unbounded tidal accelerations between neighbouring trajectories."}. So it seems that embedding the BBMB black hole in an external magnetic field emphasizes its singular behaviour. For these reasons one has to be very careful before considering the metric of the magnetised BBMB black hole (\ref{magn-BBMB}) as a truly black hole space-time, but rather it discloses naked singularities.  However the analysis of the space-time's causal structure is beyond the scope of this work and will be done elsewhere. \\ 
The introduction of the cosmological term in the action usually helps to regularise these divergences because they can be hidden behind the horizon, as it occurs in the MTZ black hole \cite{martinez}, where the scalar field is regular on (and outside) the horizon. Moreover the cosmological constant improves the astrophysical likely-hood, but unfortunately a generating technique in presence of cosmological constant (nor even a Harrison transformation) is not available at the moment. Alternatively to the cosmological constant maybe would be sufficient to consider the acceleration in order to regularise the solution. The mathematical reason is due to the same asymptotic quadratic power scaling in the radial coordinate of the metric, between the acceleration and the cosmological constant terms. In fact, as observed in \cite{greco}, the accelerating BBMB black hole has better behaved scalar field on the horizon, because instead of being singular on the whole surface horizon, it is divergent just on one point, the pole ($\rho=m/2,\ \t=\pi$). Moreover the introduction of an external electromagnetic field into these accelerating BBMB black holes recently found in \cite{greco} and \cite{andres-hideki} will allow to remove both the conical singularities typical on the poles of this accelerating solution, as firstly discovered by Ernst himself in \cite{ernst-remove} for the c-metric. These metrics, regularised by external magnetic fields, are of particular interest because they describe black hole pairs creation \cite{kastor}, \cite{hawking}. The solution generating technique developed in this paper is able to generate such that solutions in presence of a conformally coupled scalar field, work in these direction is in progress \cite{progress}. \\
We just want to remark on some features that appear in common with the non-magnetised black hole. For instance the surface gravity
$k$, defined by $ k^2=-\mezzo \nabla^\m \chi^\n \nabla_\m \chi_\n $, where $\chi^\m$ is the killing vector $\p_t$, remains null (for $\rho=m/2$) as in the case of BBMB metric. This is typical behaviour of double degenerate horizon, such as extremal black holes. \\ 
Also the topology of constant radial slices remains the same as in the non-magnetised case; in fact consider the surface $\mathcal{S}$ described by the two dimensional metric $\bar{g}_{\m\n}$ obtained by (\ref{magn-BBMB}) fixing $\rho=\bar{\rho}=$const and $t=$const. Its Euler characteristic is: 
\beq
      \chi(\mathcal{S})=\frac{1}{4\pi}  \int_{\mathcal{S}} \sqrt{\bar{g}} \ \bar{\textrm{R}} \ d\t \ d\phi \   =  2 \nn \\ 
\eeq
so, since $\chi(\mathcal{S}) = 2 - 2g$, the genus of the surface $\mathcal{S}$ is 0: spherical topology $S^2$. The area $\mathcal{A}$ of constant radial (and time) slices  is remarkably unchanged by the presence of the external magnetic field:
$$ \mathcal{A} = \int_0^{2\pi} d\phi \int_0^\pi d\t \sqrt{g_{\t\t}} \sqrt{g_{\phi\phi}} = 4\pi \bar{\rho}^2$$
Of course the geometry of constant radial slices is not spherical any more, but stretched along the direction of the external magnetic field.
 \\
Furthermore note that, even though the BBMB metric precisely coincides with the one of extremal Reissner-Nordstrom, the resulting magnetized BBMB (\ref{magn-BBMB}) differs from the magnetized Reissner-Nordstrom (which is not even static), since the generating technique is strongly theory dependent.\\

\subsection{Charged BBMB black hole in Melvin magnetic universe}
\label{Q-BBMB-magn}

The magnetised Penney solution (\ref{magn-penney}), for the minimally scalar coupling, found in section \ref{magnetic-penney}, can be uplifted as a solution of the Einstein-Maxwell with a conformally coupled scalar field by the set of transformations (\ref{conf-tras-psi})-(\ref{conf-tras-g}):
\bea
\hat{\Psi} &=& \sqrt{\frac{6}{k}} \left[ \frac{\left(\frac{R-a}{R-b}  \right)^{\sqrt{\frac{1-A^2}{3}}} -1}{\left(\frac{R-a}{R-b}  \right)^{\sqrt{\frac{1-A^2}{3}}} + 1}   \right] \label{gen-scalar} \\
\hat{ds}^2 &=& \frac{1}{4} \left[ \left(\frac{R-a}{R-b}  \right)^{\sqrt{\frac{1-A^2}{3}}}  + \left(\frac{R-a}{R-b}  \right)^{-\sqrt{\frac{1-A^2}{3}}} +2 \right] ds^2_{(magn-Penney)} \label{general}
 \eea
The electromagnetic potential $A_\m$ remains unchanged as in (\ref{A_phi}) and (\ref{A_tau}) because of the conformal invariance of the Maxwell coupling in four dimensions.\\
As summarised in the following table \ref{table1} (where also some other notable space-times are listed) this solution contains both the Ernst metrics family, such as magnetized Reissner-Nordstrom black hole ($A=1$) and the magnetized and charged BBMB metric for $A=1/2$. 

\begin{table}[h] 
 \begin{center}
 \begin{tabular}{|c|c|c|c|c|} 
  \hline
 SPACE-TIMES & $A$ & $B$ & $e_0$ & $m$ \\ 
 \hline 
 \rule[-1ex]{0pt}{2.5ex} Magnetised charged BBMB & 1/2 & $\surd$ & $\surd$ & $\surd$ \\ 
 \hline 
 \rule[-1ex]{0pt}{2.5ex} Charged BBMB & 1/2 & 0 & $\surd$ & $\surd$ \\ 
 \hline 
 \rule[-1ex]{0pt}{2.5ex} BBMB Black Hole & 1/2 & 0 & 0 & $\surd$ \\ 
 \hline 
 \rule[-1ex]{0pt}{2.5ex} Magnetised Reissner-Nordstrom  & 1 & $\surd$ & $\surd$ & $\surd$ \\ 
 \hline 
 \rule[-1ex]{0pt}{2.5ex} Reissner-Nordstrom & 1 & 0 & $\surd$ & $\surd$ \\ 
 \hline 
 \rule[-1ex]{0pt}{2.5ex} Magnetised Schwarzschild & 1 & $\surd$ & 0 & $\surd$ \\ 
 \hline 
 \rule[-1ex]{0pt}{2.5ex} Schwarzschild & 1 & 0 & 0 & $\surd$ \\ 
 \hline 
 \rule[-1ex]{0pt}{2.5ex} Melvin magnetic universe & $\forall$ & $\surd$ & 0 & 0 \\ 
 \hline 
 \rule[-1ex]{0pt}{2.5ex} Minkowski & $\forall$ & 0 & 0 & 0 \\ 
 \hline  
 \end{tabular}  
 \caption{Some specialisation of the metric (\ref{general}), for some values of its parameters.}\label{table1}
\end{center}
\end{table}

 Let's analyse this point, so henceforth $A$ will be fixed to $1/2$. Moreover we perform a change of the radial coordinate:
\beq
        R \longrightarrow \frac{4\rho^2-ab}{4\rho-a-b} 
\eeq
We prefer to express the $A=1/2$ solution just in terms of mass and charge parameters $m$ and $e_0$, instead of the less physical $a$ and $b$. They are related (setting the coupling constant ratio $G/\m_0=1$) as follows : $2m=a+b$ and  $e_0=abA$. So (\ref{gen-scalar}) and (\ref{general}) take form:
\bea
     \hat{\Psi} &=& \sqrt{\frac{6}{\kappa}} \left( \frac{\sqrt{m^2-4e_0^2}}{2\rho-m} \right) \\
     \hat{ds}^2 &=& |\Lambda|^2 \left[ - \left(  1 - \frac{m}{2\rho}  \right)^2 d\tau^2 + \frac{d\rho^2}{\left(  1 - \frac{m}{2\rho}  \right)^2} + \rho^2 d\theta^2 \right] + \frac{\rho^2 \sin^2 \theta}{|\Lambda|^2} (d\phi-\omega dt)^2   \label{magn-charg-BBMB}
\eea
 where 
\bea
      e^\beta (\rho)  \  &=& \frac{4\rho^2(\rho^2-m\rho+e_0^2)}{(2\rho-m)^2} \\
      \Lambda(\rho,\theta) &=& 1 + iBe_0\cos\t+\frac{B^2}{4} \left[e^\beta \sin^2 \t +e_0^2 \cos^2\t \right] \\
        \omega(\rho,\t) &=& \frac{B^3}{4} e_0 \sin^2\t \left( 2\rho -2m + \frac{2 e_0^2}{\rho} + \frac{m^2-4e_0^2}{2\rho-m} \right) - B^3 e_0 \frac{2\rho^2-2e_0^2}{2\rho-m} + \frac{1}{2\rho}\left(4Be_0-B^3e_0^3 \right) \nn \\
A_\phi(\rho,\t) &=& - \frac{\frac{B}{2}(e^\beta \sin^2 \t +e_0^2 \cos^2\t)[1+\frac{B^2}{4}(e^\beta \sin^2 \t +e_0^2 \cos^2\t)]+Be_0^2\cos^2\t}{|\Lambda|^2} \\
A_\tau(\rho,\t) &=& -\frac{3}{8} B^2 \sin^2 \t \left(2\rho -2m + \frac{2 e_0^2}{\rho} + \frac{m^2-4e_0^2}{2\rho-m} \right) + \frac{3 e_0 B^2 (\rho^2 -e_0^2)}{2\rho-m} + \frac{3e_0^3B^2}{4\rho} - \frac{e_0}{\rho}  \quad . \nn
\eea  

This solution describes a charged BBMB black hole embedded in an axial external magnetic field. The same considerations of the previous section \ref{magnetic-BBMB} about the appearance of curvature singularities on the poles of the surface $\rho=m/2$ have to be taken into account with caution. The immersion into a background magnetic field is, therefore, not so physically smooth as for more standard back holes, such as the Kerr-Newman family, although mathematically similar. \\
The fact that the seed black hole is charged and immersed into a external magnetic field leads to frame dragging effects, due to the $\overrightarrow{E} \times \overrightarrow{B}$ circulating momentum flux in the stress-energy tensor, which serves as a source for a twist potential. Thus, although the seed metric is static, the Harrison-transformed one is stationary. The angular momentum is proportional to the intrinsic electric charge of the black hole $e_0$ and the external magnetic field $B$, so the rotation can be detained by switching off either the black hole electric charge $e_0$, or the external magnetic field $B$. In the first case we will retrieve the static uncharged magnetised BBMB metric of subsection \ref{magn-BBMB}, while in the latter case the standard charged BBMB black hole. This is a property shared by the magnetised Reissner-Nordstrom too. \\
Another property in common with the magnetised Reissner-Nordstrom black hole is, after the process of magnetisation, the appearance of a conical singularity on the polar axis (which can be interpreted as a string with positive energy density and negative tension associated with some additional and singular stress-energy tensor on the right hand side of Einstein's equations). 
This can be seen by expanding in powers of $\t$ the $g_{\theta\theta}$ and $g_{\phi\phi}$ components of the metric (\ref{magn-charg-BBMB}) for a small circle around around $\t=0$ and $\t=\pi$.  
Eventually it is possible to avoid this extra feature, obtaining a regular space-time, just by rescaling the angular coordinate $\phi \rightsquigarrow \bar{\phi}=\phi/F$, where

\beq
 F = 1 + \frac{3}{2} B^2 e_0^2 + \frac{B^4e_0^4}{16}  \quad .
\eeq

This value is exactly in agreement with the result of \cite{hiscock} for the Reissner-Nordstrom black hole. This feature is not present (i.e. $F=1$) when the intrinsic electric charge of the black hole $e_0$ is null, as in the metric (\ref{magn-BBMB}). \\
Note that, while the static magnetised BBMB metric (\ref{magn-BBMB}) approaches the Melvin magnetic universe asymptotically, this stationary solution (\ref{magn-charg-BBMB}) does not reach globally (i.e. for all $\t$) the Melvin universe because the electric field on the symmetry axis is not null for $\rho\rightarrow\infty$, as it occurs in the magnetic universe. This is a common fact for magnetised charged black holes, as pointed out in \cite{hiscock}.\\ 
The generalisation to the dyonic black hole (that is using as a seed a metric with an intrinsic magnetic potential in addition to the electric one) is trivial because of the electromagnetic duality of the Maxwell field in four dimension.\\
In this section we proposed just a couple of examples, but note that applying the transformations $(I)-(VI)$ one can generate an infinite tower of physically inequivalent solution, exactly as happens for the Einstein-Maxwell theory.
\\

\section{Comments and Conclusions}
 
In this paper we have applied the Ernst's solution generating technique to Einstein gravity coupled with Maxwell electromagnetism and a minimally coupled scalar field. We have found that, for axisymmetric and stationary space-times, the SU(2,1) symmetry group behind the Kinnersley transformation is preserved and can be used to generate an infinite tower of solutions. A couple of examples are provided and worked out to show how the machinery works. In particular the Fisher, Janis, Robinson, Winicour metric and the Penney metrics are embedded in an external magnetic field thanks to an Harrison transformation. \\
Then these magnetised naked singularities, by means of a conformal transformation, are mapped to uncharged and charged BBMB black holes embedded in an external Melvin magnetic universe for Einstein-Maxwell theory of gravitation with a conformally coupled scalar field. The "intrinsic" charged metric is stationary rotating while the uncharged remains static after the Harrison transformation. The external magnetic field seems to sharpen the singular behaviour of the standard BBMB black holes, because singularities not covered by event horizons come out.  \\
Therefore Ernst's solution generating technique can be stretched also in presence of a conformally coupled scalar field, acting on the seed metric through a sequence of three steps: $(1)$ a conformal transformation $f$ that brings the seed metric to the minimally coupled (MC) system, $(2)$ then any (let's say $n$) sequence of generalised Kinnersley $g_1\circ g_2 \circ \dots \circ g_n$ transformations can be performed in the MC system and finally $(3)$ come back to the conformally coupled (CC) system with a conformal transformation $f^{-1}$, as is represented in following figure:
\begin{displaymath}
\xymatrix{
MC  \ar[d]_{g_1\circ g_2 \circ \dots \circ g_n} &
CC \ar[l]_f \ar[d]^{\hat{g}=f^{-1}\circ g_1\circ g_2 \circ \dots \circ g_n\circ f} \\
MC \ar[r]_{f^{-1}}
&  CC}
\end{displaymath}
 
We suspect that, similarly to what occurs in the case with a vanishing scalar field (Gerosh theorem), also for the minimally (and conformally) coupled scalar field to Einstein-Maxwell gravity $all$ space-times solutions might be generated by the set of generalised Kinnersley transformations $(I)-(VI)$.
The biggest issue we are concerned about is the suitability of the conformally rescaled Lewis-Weyl-Papapetrou metric for describing the most general stationary, axisymmetric space-times for Einsten-Maxwell theory with a conformally coupled scalar field.
 \\
It worth to noticing that in this paper we only take advance of the duality between minimally and conformally coupled scalar field but Ernst's solution generating technique here considered can be applied to many other theories connected with the minimally coupled scalar matter, such as some class of Brans-Dicke or F(R) gravities. \\ 
Furthermore the same procedure can be directly extended also to more general matter such as harmonic map coupling, consisting in a collection of scalar fields arranged in a non-linear sigma-model fashion, and all conformally related theories. Generally in that case the group of symmetry is enlarged. \\
So the mechanism developed here is able to generate an infinite number of physically inequivalent axisymmetric stationary solutions for a wide range of gravitational theories related to the scalar coupling (and eventually to Maxwell electromagnetism).  \\
For future perspective we would like to explore the possibility of exploiting integrability and the symmetries of the system directly in the conformally coupled system\footnote{This procedure can be done in principle for any theory related to Einstein-Maxwell gravity minimally coupled with a scalar field not just for a conformally coupled one.} without passing though the minimally coupled one, and try to apply this formalism also for a possible generalisation of the BBMB metrics including the Kerr family. Work in progress in this direction and in the magnetising the accelerating BBMB black hole are currently carried on. Also a better understanding of the causal structure and eventually the thermodynamic properties of magnetised BBMB space-time might be interesting. \\
For people interested in higher dimensional gravity, the generalisation to five dimension of the present work is straightforward, just following the lines of \cite{ida} and \cite{iguchi}, where the Ernst's formalism, without scalar fields, was extended to five dimensions.

\section*{Acknowledgements}
\small I would like to thank Hideki Maeda, Cristi\'{a}n Mart\'{i}nez and Patricia Ritter for fruitful discussions. \\ \small This work has been funded by the Fondecyt grant 3120236. The Centro de Estudios Cient\'{\i}ficos (CECs) is funded by the Chilean Government through the Centers of Excellence Base Financing Program of Conicyt. \\
\normalsize


\end{document}